\def\beq{\begin{equation}}
\def\eeq{\end{equation}}
\def\bea{\begin{eqnarray}}
\def\eea{\end{eqnarray}}
\def\bq{\begin{quote}}
\def\eq{\end{quote}}

\def\gappeq{\mathrel{\rlap {\raise.5ex\hbox{$>$}}
{\lower.5ex\hbox{$\sim$}}}}

\def\lappeq{\mathrel{\rlap{\raise.5ex\hbox{$<$}}
{\lower.5ex\hbox{$\sim$}}}}

\def\bbz{fa Z \kern-8.9pt Z}
\documentstyle [12pt,epsfig]{article}
\begin{document}
\thispagestyle{empty}
\vspace*{-1cm}
\begin{flushright}
{CERN-TH/97-76} \\
{hep-ph/9704339} \\
\end{flushright}
\vspace{1cm}
\begin{center}
{\large {\bf $R$-Parity Violation and Unification}} \\
\vspace{.2cm}                                                  
\end{center}

\begin{center}
\vspace{.3cm}
{\bf 
G.F. Giudice\footnote{On leave of absence from INFN, Sez. di Padova, Italy.}}
 and                                                     
{\bf R. Rattazzi}
\\                          
\vspace{.5cm}
{ Theory Division, CERN, CH-1211, Gen\`eve 23, Switzerland} \\
\end{center}                              

\vspace{2cm}

\begin{abstract}
The reported anomaly in deep-inelastic scattering at HERA has revived 
interest in the phenomenology of $R$-parity violation. From the theoretical
point of view, the existence of $R$-violating interactions poses two
considerable problems. The first one concerns the flavour structure of
the interactions and the origin of an appropriate suppression of
flavour-changing neutral-current processes and lepton-family transitions.
The second one concerns the way of embedding $R$-violating interactions
in a grand unified theory (GUT) without introducing unacceptable nucleon
decay rates. We show that the second problem can be solved by a mechanism 
which is purely group theoretical and does not rely on details of the 
flavour theory. We construct explicit GUT models in which our mechanism 
can be realized.

\end{abstract}
\vfill 
\begin{flushleft}
CERN-TH/97-76\\
April 1997\\
\end{flushleft}

\newpage

\noindent{\bf 1.} ~~The anomaly in 
deep-inelastic $e^+ p$ scattering events reported by 
H1~\cite{h1} and ZEUS~\cite{zeus} and the excess of four-jet events
observed by ALEPH~\cite{aleph}, but not confirmed by the other LEP
experiments~\cite{aleph2}, have revived interest~\cite{int,fou} in
the phenomenology of
$R$-parity violating interactions. Certainly more statistics is
required to understand if experiments are really observing some
signal of new physics. Nevertheless we believe it is timely and important
to investigate what are the consequences of $R$-parity violation in our
understanding of the theoretical framework of supersymmetric models.

The two main problems arising from $R$-parity violation are connected 
with {\it flavour} and with {\it unification}. As 
we emphasize below, we believe
that these two problems are quite different both in their quantitative
aspect (the unification problem being numerically more acute) and
in their conceptual aspect. In this paper we will present a solution
to the unification problem which is independent of the flavour
structure of the theory, and relies only on GUT symmetry properties. 
At present, the question of flavour remains unresolved, since we do not
know any fully convincing
theory which generates
the peculiar hierarchy of the Yukawa couplings -- let alone the $R$-parity
violating couplings. We believe that separating the two puzzles, and 
solving the unification problem in terms of GUTs, leads to important
theoretical progress since, as we argue below, it is probably unrealistic
to hope that an ultimate flavour theory can provide the right cure.

\noindent{\bf 2.} ~~We concentrate on the $R$-violating interaction
suggested to explain the large-$Q^2$ data at HERA, which is
given by the term in the superpotential
\beq
W=\lambda_{ijk}Q_L^i {\bar D}_R^j L_L^k 
\label{qdl}
\eeq
with $\lambda_{i11}\gappeq 4\times 10^{-2}$~\cite{int}
and $i$ equal to 2 or 3. Here
$i,j,k$ refer to generation indices and we employ a standard notation for
quark and lepton superfields. The {\it flavour} problem arises because
the generation structure of the operator in eq.~(\ref{qdl}) is in general
not aligned with the generation structure of the Yukawa interactions
\beq
W =  h^d_{ij}Q_L^i {\bar D}_R^j {\bar H} +
       h^u_{ij}Q_L^i {\bar U}_R^j { H} +
       h^e_{ij}L_L^i {\bar E}_R^j {\bar H}.
\label{yuk}
\eeq
We work in a basis where $h^d$ and $h^e$ are diagonal, and $h^u$ is a diagonal
matrix times the Kobayashi-Maskawa matrix.
Because of the mismatch in flavour space,
squarks and sleptons mediate effective four-fermion interactions
which lead to flavour-changing neutral-current processes and lepton-family
transitions. For instance, measurements of the
$K^0-{\bar K}^0$ 
mixing parameters $\Delta m_K$ and $\epsilon$
imply~\cite{bar}
\beq
|\lambda_{121}|<\left( \frac{4\times 10^{-2}}{|\lambda_{211}|}\right)~
\left( \frac{m_{\tilde \nu}}{200~{\rm GeV}}\right)^2 ~10^{-7}~,
\label{kk}
\eeq
\beq
|\lambda_{121}|<\left( \frac{4\times 10^{-2}}{|\lambda_{211}|}\right)~
\left( \frac{m_{\tilde \nu}}{200~{\rm GeV}}\right)^2 ~\frac{7
\times 10^{-10}}{\sin \delta_\lambda}~,
\eeq
where $\delta_\lambda$ is the relative phase between the two $\lambda$
couplings.
Bounds on $\mu$--$e$ conversion imply~\cite{bar}
\beq
|\lambda_{i12}|<\left( \frac{4\times 10^{-2}}{|\lambda_{i11}|}\right)~
\left( \frac{m_{{\tilde u}_i}}{200~{\rm GeV}}\right)^2 ~5\times 10^{-6}~,
\label{me}
\eeq
for any $i=1,2,3$.
While limits on a single $\lambda_{ijk}$ coupling are weak enough~\cite{bgh} 
to allow for an important phenomenological r\^ole of $R$-parity breaking,
the product of two $\lambda$ couplings with different 
generation indices is severely constrained. A successful theory of
flavour and $R$-parity violation should explain the origin of this strong
hierarchy.

Let us now turn to the {\it unification} problem. If the interaction of
eq.~(\ref{qdl}) has to be embedded in a trilinear term arising from a GUT,
then the superpotential in general also contains the interactions
\beq
W=       \lambda^\prime_{ijk}{\bar E}_R^i  L_L^j L_L^k +
\lambda^{\prime \prime}_{ijk}{\bar U}_R^i {\bar D}_R^j  {\bar D}_R^k  ~.
\label{supe}
\eeq
While the $\lambda$ and $\lambda^\prime$ couplings violate lepton number,
$\lambda^{\prime \prime}$ violates baryon number. Their simultaneous
presence is therefore strongly constrained by nucleon-decay searches.
For instance the experimental bound on $n\to K^+ e^-$ implies
\beq
|\lambda^{\prime \prime}_{i12}|\lappeq \left( \frac{4\times 
10^{-2}}{|\lambda_{i11}|}\right)~
\left( \frac{\tilde m}{200~{\rm GeV}}\right)^3 ~
\left(\frac{175~{\rm GeV}}{m_{u_i}}\right)~
 10^{-25}~.
\label{bound}
\eeq
Here $\tilde m$ is the typical supersymmetry-breaking mass 
parameter in the ${\tilde
u}_i$ mass matrix.
The presence of the quark mass $m_{u_i}$ in eq.~(\ref{bound}) is a product of
the left-right squark mixing necessary to construct the $\Delta B=
-\Delta L=-1$ four-fermion operator. Of course, in the case $i=3$, it does not
amount to any significant suppression.

{}From 
eq.~(\ref{bound}) we see that the
unification problem (or, in other words, the simultaneous presence of
baryon and lepton number violation) poses a more severe difficulty than
flavour. There can be hope that
hierarchies between couplings analogous to those required by 
eqs.~(\ref{kk})--(\ref{me}) 
can be explained in a complete theory of flavour. On the other 
hand, we prefer to believe that the observed
suppression
of nucleon decay is caused by the small ratio between the weak
and the GUT scale rather than by some broken flavour symmetry,
which generates hierarchies
as a power expansion of some parameter like the Cabibbo angle. Examples
of theories in which the baryon-number and $R$-violating interactions are
suppressed by flavour symmetries
exist~\cite{idi}, but are not embedded in a GUT. It is not 
clear how they can be unified and made consistent
with the size of couplings suggested by the HERA data. For this reason,
we believe that the solution should lie within the GUT dynamics.

\noindent{\bf 3.} ~~We now want to embed the $R$-parity violating interaction
of eq.~(\ref{qdl}) in a GUT, without running into the problem of
baryon-number violation. To start, we choose the simplest example
of $SU(5)$ and denote the matter content by ${\bf 10}^i +{\bf {\bar 5}}^i$
($i=1,2,3$) and the Higgs superfields by $H$ and $\bar H$, respectively
a ${\bf 5}$ and ${\bf {\bar 5}}$ of $SU(5)$. The Yukawa
couplings are
\beq
W=h_{ij}(\Sigma ) {\bf 10}^i {\bf 10}^j H +
{\bar h}_{ij}(\Sigma ) {\bf 10}^i {\bf {\bar 5}}^j {\bar H }~.
\label{yukg}
\eeq
Here $h_{ij}$ and ${\bar h}_{ij}$ are functions of the adjoint field $\Sigma$,
which spontaneously breaks the $SU(5)$ symmetry. 
After $\Sigma$ gets its
vacuum expectation value (VEV), 
they reproduce the
ordinary Yukawa couplings $h^u_{ij}$, $h^d_{ij}$, $h^e_{ij}$ at low energy.

The first attempt one can try is to include in the superpotential only the 
bilinear term\footnote{The phenomenology of the non-GUT version of this term
has been considered, for instance, in ref.~\cite{bil}.}
\beq
W=\rho_i {\bf {\bar 5}}^i H~,
\label{rho}
\eeq
where $\rho_i$ are mass parameters smaller than the weak mass scale. 
The operators in eq.~(\ref{rho}) could be generated by some mechanism similar
to the one responsible for the Higgs-mixing $\mu$ term. By defining appropriate
mass eigenstates, we can rotate the term in eq.~(\ref{rho}) into some
$R$-violating trilinear couplings. The ${\bar D}_R^i$ states mix with the
Higgs triplet contained in $\bar 
H$ and give rise to the baryon-number violating
coupling of eq.~(\ref{supe}) with
\beq
\lambda_{ijk}^{\prime \prime}= {\bar h}_{ij} \frac{\rho_k}{M_H}~.
\eeq
Since the Higgs-triplet mass $M_H$ is of the order of the GUT scale, we 
obtain a considerable suppression of the baryon-violating interaction.
An even further suppression exists in models where the doublet-triplet
splitting is obtained without a direct mass term $H \bar{H}$.
The mixing between the lepton superfields $L_L^i$ and the Higgs doublet
generates couplings $\lambda$ and $\lambda^\prime$ which are suppressed
only by the ratio $\rho_i/\mu$, where $\mu$ is the Higgs-mixing term of the
order of the weak scale. However, in this case, 
$\lambda_{ijk}\propto h^d_{ij}\rho_k$ and the value of the $R$-violating
coupling constant
suggested
by the HERA data is incompatible with the limit on the electron
neutrino mass which implies~\cite{int} 
\beq
|\lambda_{331}|<5\times 10^{-3}~\left( \frac{m_{\tilde b}}{200~{\rm GeV}}
\right)^{\frac{1}{2}}~.
\eeq

An interesting possibility, which was first suggested in ref.~\cite{bar}, 
is that the only $R$-parity violation comes from an operator
in the superpotential
\beq
\lambda^G_{ijk}(\Sigma ) {\bf 10}^i {\bf {\bar 5}}^j {\bf {\bar 5}}^k  ~.
\label{supeg}
\eeq
If $
\lambda^G_{ijk}(\langle \Sigma \rangle )$ and ${\bar h}_{jk}(
\langle \Sigma \rangle )$ (the Yukawa coupling 
defined in eq.~(\ref{yukg}))
are simultaneously diagonal in 
$j,k$ for any $i$ and for any $SU(5)$ index, then the 
interaction (\ref{supeg}) generates nonvanishing $\lambda$, while
$\lambda^{\prime}$ and
$\lambda^{\prime \prime}$ identically vanish \cite{bar}. 
This is simply because $\lambda_{ijk}^\prime$ and 
$\lambda_{ijk}^{\prime \prime}$ 
are antisymmetric in $j,k$,
while $\lambda_{ijk}$ has no symmetry
properties.
The coupling constants $\lambda_{ijk}^\prime$
vanish at the GUT scale, but small values are generated by the renormalization
to the weak scale.
It was also shown in ref.~\cite{bar} that the
renormalization to the weak scale gives only small further violation of
flavour in the $\lambda$ sector, 
and therefore the {\it ansatz} on the generation structure of
$\lambda_{ijk}^G$ specified above can render the $R$-breaking 
interpretation of the
HERA data compatible with unification. However this {\it ansatz}
may seem rather {\it ad hoc}. Also it seems to rely on flavour
properties, which we find a disturbing aspect, as previously discussed.

We turn now to discuss how, with no reference to the flavour theory,
a GUT can lead to $R$-parity violating couplings relevant for squark
production at HERA together with
vanishing
baryon-number violating couplings. 
In terms of GUT representations, the $R$-violating interactions
in eqs.~(\ref{qdl}) and (\ref{supe}) are written as\footnote{This defines a
complete set of operators. This can be understood by counting the
number of gauge invariants. For $N$ matter generations, the operators listed in 
eqs.~(\ref{s1})--(\ref{s4})  contain $N^2(N-1)/2$ or $N^2(N+1)/2$
flavour components, if the two ${\bf \bar 5}$ are combined in a ${\bf
\bar{10}}$ or ${\bf
\bar{15}}$, respectively. This makes a total of $N^2(2N-1)$ invariants
and matches the number of invariants of the low-energy theory, which
are described by $N^3$ couplings $\lambda$  and $N^2(N-1)/2$ couplings
$\lambda^\prime$
and $\lambda^{\prime \prime}$.}
\bea
{\cal O}_1^{ijk}\equiv {\bf {\bar 5}}^j\cdot {\bf {\bar 5}}^k\cdot {\bf 10}^i
~~~&\to &~~~\lambda_{ijk} ,~\lambda^\prime_{ijk}  , ~\lambda^{\prime 
\prime}_{ijk} \label{s1}\\
{\cal O}_2^{ijk}\equiv  
({\bf {\bar{5}}}^j \cdot  {\bf {\bar{5}}}^k)_{{\bf {\bar{10}}}}\cdot  
({\bf 10}^i\cdot  \Sigma )_{\bf 10}
~~~&\to &~~~\lambda_{ijk}  ,~\lambda^\prime_{ijk}  , 
~\lambda^{\prime \prime}_{ijk} \label{s2}\\
{\cal O}_3^{ijk}\equiv ({\bf {\bar{5}}}^j\cdot   {\bf 
{\bar{5}}}^k)_{{\bf {\bar{15}}}}\cdot  
({\bf 10}^i\cdot  \Sigma )_{\bf 15}
~~~&\to &~~~\lambda_{ijk}  \label{s3}\\
{\cal O}_4^{ijk}\equiv \left(({\bf {\bar{5}}}^j \cdot  
{\bf {\bar{5}}}^k)_{{\bf {\bar{10}}}}
\cdot \Sigma \right)_{{\bf {\bar{15}}}}
\cdot  
({\bf 10}^i\cdot  \Sigma )_{\bf 15}
~~~&\to &~~~\lambda_{ijk} ~.\label{s4}
\eea
Here we have specified the contractions of the $SU(5)$ indices with a pendix
denoting the product representation. 
Operators with more powers of $\langle \Sigma \rangle$ reduce to combinations
of the above since, for any $n$, $\langle \Sigma \rangle^n$ 
is a linear combination of $\langle \Sigma \rangle$ and the identity.
We have also marked explicitly which
of the couplings $\lambda$, $\lambda^\prime$, or $\lambda^{\prime \prime}$
are generated by the various operators after $\Sigma$ gets its VEV. If we
select the operators ${\cal O}_3$ or ${\cal O}_4$, at low energies we retain
only the $\lambda$ couplings. This 
can be easily implemented in a GUT, since specific operators
can be selected by appropriately choosing the virtual states
which generate them. This selection is a consequence of group-theoretical
properties and does not rely on the flavour dynamics.

${\cal O}_3^{ijk}$ combines
the two ${\bf \bar 5}$ in a symmetric state. 
Therefore it is symmetric
in the indices $j$ and $k$ and its contribution to $\lambda^\prime$
and $\lambda^{\prime \prime}$ identically vanishes.
We will give later an example of how this
case can be realized in GUTs. ${\cal O}_4^{ijk}$  also selects just the
$\lambda$ coupling, although it is antisymmetric in the flavour
indices $j$ and $k$. Only the coupling $\lambda$ survives because
$({\bf 10}^i\cdot  \Sigma )_{\bf 15}$ is projected only onto $Q_L^i$, after
GUT symmetry breaking.

Solving the problem of baryon-number and $R$-parity violating interactions
with GUT dynamics may not be sufficient. We are now concerned with 
higher-dimensional operators suppressed by powers of the Planck mass $M_{P}$
generated by the unknown dynamics of quantum gravity. These operators
may have the most generic structure compatible with the unbroken
symmetries, and therefore reintroduce the unwanted $\lambda^{\prime
\prime}$ couplings in the low-energy effective theory. These couplings
will only be suppressed by some powers of $M_{GUT}/M_{P}$, and therefore
bounds like the one in eq.~(\ref{bound}) require that these operators
should not be present at least up to some high dimensionality.

The non-renormalization theorems~\cite{nor} of supersymmetry can protect the
GUT theory from this danger. If the operator ${\cal O}_3$ or 
${\cal O}_4$ is generated
only after some stage of symmetry breaking, this same 
symmetry together with the constraint
of holomorphicity of the superpotential
can forbid any dangerous higher-dimensional operator. This mechanism, in
which a broken symmetry protects against the appearence of certain
terms in the superpotential at all orders, has also been used in the
constructions of flavour theories~\cite{fla}. Of course the renormalization
of the K\"ahler function is not under control and it
can effectively generate new terms in the 
superpotential, once supersymmetry is broken. The size of these effects,
which are proportional to some power of the ratio between the 
supersymmetry-breaking scale and $M_{P}$ can be estimated in a given
model and then compared with the experimental bound on nucleon decay.

\noindent{\bf 4.} ~~Our mechanism is best illustrated by a simple example
in $SU(5)$. To generate the desired operators
${\cal O}_3$ and ${\cal O}_4$, we introduce 
some fields in the symmetric product
of two fundamentals, $S+\bar{S}$, which transform as ${\bf 15} +{\bf {\bar 
{15}}}$. The presence of the field $S$ together with its conjugate $\bar S$
insures the cancellation of $SU(5)$ anomalies and allows a superheavy
mass term. Let us consider the following interaction for
the fields $S$ and $\bar S$:
\beq
W= {\bf \bar 5}^i {\bf \bar 5}^j S + {\bar S} {\bf 10}^i \Sigma + S {\bar S}
\phi ~.
\label{ss}
\eeq
Here $\phi$ is a gauge singlet, which plays the r\^ole of a mass parameter,
and $\langle \phi \rangle \equiv M_X$ is somewhat larger than the
GUT scale. The effective theory below $M_X$, obtained by integrating out 
$S$ and $\bar S$, contains the operator ${\cal O}_3$.
Just below $M_{GUT}$, 
$\lambda_{ijk}$ is generated, but $\lambda^\prime_{ijk}$ and
$\lambda^{\prime \prime}_{ijk}$ vanish.

In order to explain the desired structure of $R$-breaking interaction, we
also have to justify the
absence of
the renormalizable coupling ${\bf 10}^i {\bf 5}^j {\bf 5}^k$. As explained 
above, this is in general not sufficient, because the strong bounds on
nucleon decay also require the absence of a large number of higher-dimensional
operators. The simplest symmetry we can introduce is an abelian 
{\it flavour-independent} $U(1)$.
The $U(1)$ charge assignment is completely determined by the requirement
that the most general superpotential consistent with the $SU(5)\times
U(1)$ symmetry is given by eqs.~(\ref{yukg}) and 
(\ref{ss}) together with terms responsible for the GUT symmetry breaking
involving $\Sigma$, bilinears in ${\bar H} H$, and possibly other fields.
We find the following $U(1)$ charges: $X({\bf 10}^i)=-1$, $X({\bf 
{\bar 5}}^i)=3$, $X(H)=2$, $X({\bar H})=-2$, $X(\Sigma )=0$, 
$X(S)=-6$, $X({\bar S})=1$, $X(\phi )=5$. 
This $U(1)$ is anomalous, but the Green-Schwarz mechanism~\cite{gre} 
can be invoked to cancel the gauge anomalies. It is interesting that 
an anomalous $U(1)$ group usually appears in the effective field 
theory derived from strings~\cite{dix}. The effective theory then contains a
Fayet-Iliopoulos term, equal to~\cite{dix}
\beq
\xi =\frac{g^2{\rm Tr}~X}{192 \pi^2}M_P^2~.
\eeq
If the signs of $X(\phi )$ and
${\rm Tr} X$ are opposite 
(and, in our example, this is true at least in the observable sector), then
$\phi$ can get a VEV, given by
\beq
\langle \phi \rangle \equiv M_X = \sqrt{\frac{-\xi}{X(\phi )}} 
~.
\eeq

The theory has an accidental discrete symmetry, under which ${\bf 10}^i$,
${\bf {\bar 5}}^i$, $\bar S$, and $\phi$ are odd, while all other 
chiral superfields are even. This can be identified with the usual $R$
parity, and it is broken by $\langle \phi \rangle$ at the scale
$M_X$. The size of the
low-energy $R$-parity violating coupling $\lambda$ is 
$\lambda \sim {\cal O}(M_{GUT}/M_X)$.

We turn now to discuss the suppression of higher-dimensional operators.
The property of holomorphicity of the superpotential and the $U(1)$ symmetry
forbid all possible quantum-gravity derived operators, which give rise
to $R$-parity breaking in the low-energy effective theory. Indeed 
terms
of the generic form
\beq
\int d^2\theta H{\bf {\bar 5}}^i f(\Sigma ,\phi ) ~~~{\rm or}~~~
\int d^2\theta {\bf 10}^i
{\bf {\bar 5}}^j {\bf {\bar 5}}^k g(\Sigma ,\phi )
\eeq
cannot appear since holomorphicity requires only positive powers of $\phi$
in the functions $f$ and $g$, while $U(1)$ invariance requires a
negative power of $\phi$. Of course this property depends on
the particular charge assignment and it would not hold if there existed
fields which acquire VEVs of the order of the GUT scale and have
negative $U(1)$ charges. 

Planck-mass suppressed operators can also affect the dynamics of the fields
$S$ and $\bar S$. In particular the most general interactions consistent
with $SU(5)\times U(1)$ symmetry have 
the same form of
those in eq.~(\ref{ss}), with arbitrary insertions of $\Sigma$ fields.
The key point is that these operators are 
not going to modify our mechanism, since 
they can generate ${\cal O}_4$, but never ${\cal O}_1$ or ${\cal O}_3$.
We can understand this result differently by considering the
$SU(3)\times SU(2) \times U(1)$ content of the {\bf 15}. 
The {\bf 15} does not contain any standard model representation with the
correct quantum numbers to mediate effective interactions ${\bar U}_R
{\bar D}_R {\bar D}_R$ or ${\bar E}_R L_L L_L$.
Only a colour triplet, weak doublet can be propagated between
the ${\bf \bar 5}^i {\bf \bar 5}^j$ and ${\bf 10}^k \Sigma$ states.
From this point of view, our mechanism is analogous to the
missing partner
mechanism~\cite{mis} used to split the masses of the Higgs doublet and triplet
belonging to the same GUT representation\footnote{The use of an anomalous
$U(1)$ to implement to all orders the missing partner mechanism was studied
in ref.~\cite{nulla}.}.

The K\"ahler function can contain terms of the kind
\beq
\int d^4\theta H{\bf {\bar 5}}^i \phi^\dagger F(Z,Z^\dagger ) ~~~{\rm and}~~~
\int d^4\theta {\bf 10}^i
{\bf {\bar 5}}^j {\bf {\bar 5}}^k \phi^\dagger G(Z,Z^\dagger )~,
\label{kah}
\eeq
where $Z$ is the spurion superfield which parametrizes 
supersymmetry breaking and has a non-zero VEV of the
auxiliary field. The terms in eq.~(\ref{kah})
give rise to the low-energy parameters $\rho \sim (m_{\tilde G} M_X)/M_{P}$
and $\lambda \sim \lambda^\prime \sim \lambda^{\prime \prime}
\sim (m_{\tilde G} M_X)/M^2_{P}$, where $m_{\tilde G}$ is the gravitino
mass. In theories where the breaking of supersymmetry is communicated
to the observable sector
by gravity~\cite{grr} $m_{\tilde G}$ is of the order of the weak scale.
Then, a comparison with the bound in eq.~(\ref{bound}) shows that a
certain degree of suppression coming either from the flavour theory or
{}from quantum gravity is necessary. Lacking much knowledge of either of
the two theories, we cannot exclude this case. On the other hand, in theories
where supersymmetry breaking is communicated by particles lighter than
$M_P$, $m_{\tilde G}$ is smaller than the weak scale and it can efficiently
suppress any K\"ahler-induced $R$-parity violation.

The last comment we wish to make about this model concerns
the flavour structure of the $R$-parity violation. Although our mechanism
does not address the question of flavour, as a byproduct of this model,
we obtain a restriction on the generation structure of the $\lambda_{ijk}$
couplings which, just below the GUT scale, have the form
\beq
\lambda_{ijk} = A_i B_{jk}~.
\label{fac}
\eeq
This is a consequence of our assumption that a single state $S+{\bar S}$
mediates the effective interaction which generates the operator
${\cal O}_3$.
Only within a  complete theory of flavour can we hope to understand the
hierarchical structure of the vector $A$ and the matrix $B$.

\noindent{\bf 5.} ~~The model discussed in the previous section is certainly
not the only possibility to realize our mechanism. Instead of an anomalous
$U(1)$ symmetry, one could use an $R$ symmetry which easily protects against
higher-dimensional terms. However, in this case, one should specify the whole
model, including the GUT symmetry breaking sector, which it was left 
undetermined in our previous 
example.

Another possibility is to consider different GUT groups, 
flipped $SU(5)$~\cite{fli} being a very interesting option. The GUT group is
$SU(5)\times U(1)$, with matter transforming as $({\bf 10},-1)+({\bf \bar 5}
,3)+({\bf 1},-5)$, and the usual Higgs doublets embedded in $H=({\bf 5},2)$
and ${\bar H}=({\bf \bar 5},-2)$. The Yukawa couplings are
\beq
W=h^u_{ij} {\bf 10}^i {\bf \bar 5}^j {\bar H} +
h^d_{ij} {\bf 10}^i {\bf 10}^j { H } +
h^e_{ij} {\bf \bar 5}^i {\bf 1}^j { H }~,
\label{yukf}
\eeq
where we have denoted the matter superfields by their $SU(5)$ content.
The GUT symmetry breaking is triggered
by the VEVs of the fields $K=({\bf 10},-1)$
and ${\bar K}=({\bf \bar {10}},1)$, which allow a simple implementation
of the missing-partner mechanism~\cite{fli} through the interactions
\beq
W = HKK +{\bar H}{\bar K}{\bar K} ~.
\label{dou}
\eeq

An interesting feature of flipped $SU(5)$ is that renormalizable $R$-parity 
interactions are forbidden by gauge invariance. 
The low-energy $R$-violating operators in eqs.~(\ref{qdl}) and (\ref{supe}) 
are generated by the following GUT operators:
\bea
O_1^{ijk}\equiv ({\bf 10}^j \cdot  {\bf 10}^k)_{{\bf {\bar{5}}}}\cdot  
({\bf \bar 5}^i\cdot  K )_{\bf 5}
~~~&\to &~~~\lambda_{ijk}  \label{buono} \\
O_2^{ijk}\equiv ({\bf \bar 5}^i \cdot  {\bf 10}^j)_{{\bf 5}}\cdot  
({\bf 10}^k  \cdot K )_{\bf \bar 5}
~~~&\to &~~~\lambda_{ijk} , ~\lambda^{\prime \prime}_{ijk}\\
O_3^{ijk}\equiv \left( ({\bf 10}^j \cdot  {\bf 10}^k)_{{\bf {\bar{45}}}}\cdot
(K\cdot {\bar K})_{\bf 24} \right)_{\bf \bar 5}\cdot  
({\bf \bar 5}^i\cdot  K )_{\bf 5}
~~~&\to &~~~\lambda_{ijk}  \\
O_4^{ijk}\equiv {\bf \bar 5}^j \cdot  {\bf \bar 5}^k \cdot  {\bf 1}^i \cdot  K
~~~&\to &~~~\lambda^\prime_{ijk} ~.
\eea
This is a complete set of operators. Of course one can rearrange the
contractions of the $SU(5)$ indices or take linear combination of the
various operators. By doing this, we can construct an operator which
selects only the $\lambda^{\prime \prime}$ couplings:
\beq
O_5^{ijk}\equiv \left( ({\bf 10}^j \cdot  K)_{{\bf {\bar{5}}}}\cdot
({\bf 10}^k \cdot  K)_{\bf \bar 5} \right)_{\bf \bar {10}}\cdot  
({\bf \bar 5}^i\cdot  {\bar K} )_{\bf 10}
~~~\to ~~~\lambda_{ijk}^{\prime \prime}
\eeq
This operator is generated by exchange of states with flipped $SU(5)$ quantum
numbers $({\bf 10},4)$, $({\bf \bar{10}},-4)$.

If we select only $O_1$ or $O_3$, we obtain the $R$-violating coupling
invoked for an interpretation of the HERA data, but forbid the baryon-number
violating ones. The operator $O_1$ is generated by the virtual exchange
of the superfields $S+{\bar S}$, transforming as
$({\bf 5},2)+
({\bf \bar 5},-2)$, with interactions
\beq
W= {\bf 10}^i {\bf 10}^j S + {\bar S} {\bf {\bar 5}}^i K + S{\bar S} \phi ~,
\label{tut}
\eeq
where $\phi$ is a gauge singlet such that $\langle \phi \rangle =M_X$. 
However we have to forbid the couplings ${\bf 10}^iKS+\bar S {\bf \bar 5}^i
{\bf 10}^j$, which generate the operator $O_2$ and introduce the
baryon-number violating couplings $\lambda^{\prime \prime}$. All unwanted
structures are eliminated in renormalizable and non-renormalizable 
interactions by a discrete $R$-parity and an anomalous $U(1)$. The superfields
${\bf 10}^i$, ${\bf \bar 5}^i$, ${\bf 1}^i$, $\bar S$, and $\phi$ are odd
under $R$-parity , while all others are even.
The charge assignment of the anomalous $U(1)$, up to a linear combination 
with the $U(1)$ of flipped $SU(5)$, is 
$X({\bf {\bar 5}}^i)= -X({\bf 1}^i)=-X({\bar H})=2X({\bar K})=
-X({\bar S})=X(\phi )=1$, while all other fields are neutral.
Both the $R$-parity and the anomalous $U(1)$ 
are spontaneously broken the VEVs of $\phi$ and $\bar K$.
There are no fields with GUT scale VEV and negative charge $X$, and therefore
holomorphicity and symmetry invariance protect against operators with
unwanted structures of $R$-parity breaking.
As in the previous example, higher-dimensional operators involving the
fields $S$ and $\bar S$ do not affect the mechanism, since only a single
component of the $S$ field (a colour triplet, weak doublet) is propagated
in the exchange between the ${\bf 10}^i {\bf 10}^j$ and ${\bf \bar 5}^k
K$ states. We also notice that, in the flipped $SU(5)$ case, the flavour
structure of the $R$-violating coupling $\lambda_{ijk}$ factorizes between
quark and lepton indices
\beq
\lambda_{ijk} = A_{ij} B_k~,
\eeq
in contrast to the case of eq.~(\ref{fac}).

Finally notice that the operator $O_4$ can be generated by $S$-exchange
through the couplings ${\bf \bar 5}^iK\bar S +S{\bf \bar 5}^i {\bf 1}^j$.
The simultaneous presence of $\lambda$ and $\lambda^\prime$ interactions
is constrained by flavour and lepton violating processes.
However, since flipped $SU(5)$ is not based on a simple group and the
unification is not complete, the couplings $\lambda$ and $\lambda^\prime$
are here unrelated. It is therefore possible that some flavour symmetry
suppresses $\lambda^\prime$ without affecting the coupling $\lambda$
invoked to explain the HERA anomaly.

In the case of $SO(10)$, left-right symmetry forbids $R$-violating
dimension-four operators. We can then consider dimension-five 
operators of the kind 
${\bf 16}^i {\bf 16}^j {\bf 16}^k {\bf 16}_H$, where ${\bf 16}^i$
describes a generation of matter and ${\bf 16}_H$ is the Higgs
representation which breaks $SO(10)$ into $SU(5)$. Since these operators
are $SU(5)$ invariant, they predict $\lambda =\lambda^\prime =
\lambda^{\prime \prime}$. It is possible to construct higher-dimensional
operators which select only the $\lambda$ coupling. For instance
\beq
({\bf 16}^j\cdot {\bf 16}_H )_{\bf 10} \cdot
({\bf 16}^k\cdot {\bf 16}_H )_{\bf 10} \cdot {\bf 54} ~~~{\rm and}~~~
{\bf 16}^i\cdot ({\bf 45} \cdot {\bf 54} )_{\bf 45} {\bf \bar {16}}_H ~,
\eeq
which are mediated by a heavy ${\bf 45}$ and ${\bf 10}$ respectively, 
can give the right structure of $R$-violation, once the heavy ${\bf 54}$
is exchanged. The model is obviously rather involved and depends on the
unspecified dynamics which characterize the symmetry-breaking pattern.

\noindent{\bf 6.} ~~We have shown that the $R$-parity violating 
interaction suggested by the HERA data and the observed absence of
fast nucleon decay is compatible with the idea of grand unification.
The interactions between matter and the heavy fields which break the
GUT symmetry can split the different $R$-violating interactions.
By choosing
an appropriate field content, it is possible to select a preferred
pattern of $R$-parity breaking. 
This pattern is protected against higher-dimensional operators with
different $R$-violating structures by the combined effect of a symmetry,
broken at the GUT scale, and the holomorphicity of the superpotential.

Our mechanism relies on purely
group-theoretical properties and does not require any assumption about
flavour structure. We have shown specific examples of GUT models in
which our idea can be realized. Only in the context of a complete
theory of flavour will it be possible to address the question of 
flavour-changing
neutral current processes and lepton-family transitions.

We wish to thank G. Altarelli, G. Dvali, and M. Mangano for useful
discussions.

\def\ijmp#1#2#3{{\it Int. Jour. Mod. Phys. }{\bf #1~}(19#2)~#3}
\def\pl#1#2#3{{\it Phys. Lett. }{\bf B#1~}(19#2)~#3}
\def\zp#1#2#3{{\it Z. Phys. }{\bf C#1~}(19#2)~#3}
\def\prl#1#2#3{{\it Phys. Rev. Lett. }{\bf #1~}(19#2)~#3}
\def\rmp#1#2#3{{\it Rev. Mod. Phys. }{\bf #1~}(19#2)~#3}
\def\prep#1#2#3{{\it Phys. Rep. }{\bf #1~}(19#2)~#3}
\def\pr#1#2#3{{\it Phys. Rev. }{\bf D#1~}(19#2)~#3}
\def\np#1#2#3{{\it Nucl. Phys. }{\bf B#1~}(19#2)~#3}
\def\mpl#1#2#3{{\it Mod. Phys. Lett. }{\bf #1~}(19#2)~#3}
\def\arnps#1#2#3{{\it Annu. Rev. Nucl. Part. Sci. }{\bf
#1~}(19#2)~#3}
\def\sjnp#1#2#3{{\it Sov. J. Nucl. Phys. }{\bf #1~}(19#2)~#3}
\def\jetp#1#2#3{{\it JETP Lett. }{\bf #1~}(19#2)~#3}
\def\app#1#2#3{{\it Acta Phys. Polon. }{\bf #1~}(19#2)~#3}
\def\rnc#1#2#3{{\it Riv. Nuovo Cim. }{\bf #1~}(19#2)~#3}
\def\ap#1#2#3{{\it Ann. Phys. }{\bf #1~}(19#2)~#3}
\def\ptp#1#2#3{{\it Prog. Theor. Phys. }{\bf #1~}(19#2)~#3}

\end{document}